\documentstyle[preprint,pra,aps]{revtex}

\begin{document}

\bibliographystyle{prsty}
\draft
\tighten

\title{Nonclassical correlations of phase noise and photon number
in quantum nondemolition measurements}

\author{Holger F. Hofmann}
\address{Department of Physics, Faculty of Science, University of Tokyo\\
7-3-1 Hongo, Bunkyo-ku, Tokyo113-0033, Japan}


\date{\today}

\maketitle

\begin{abstract}
The continuous transition from a low resolution quantum nondemolition
measurement of light field intensity to a precise measurement of photon 
number is described using a generalized measurement postulate. 
In the intermediate regime, quantization appears as a weak modulation
of measurement probability. In this regime, the measurement result
is strongly correlated with the amount of phase decoherence 
introduced by the measurement interaction. In particular,
the accidental observation of half integer photon numbers preserves 
phase coherence in the light field, while the accidental observation 
of quantized values increases decoherence.
The quantum mechanical nature of this correlation is discussed
and the implications for the general interpretation of quantization
are considered. 
\end{abstract}
\pacs{PACS numbers:
03.65.Bz  
42.50.Dv  
42.50.Lc  
}

\section{Introduction}
In classical measurements, infinite precision is always desireable.
Therefore there is no need for a fundamental measurement theory describing 
limited resolution. Instead, the lack of precision in any actual
measurement is either neglected or considered to be an error which 
degrades the value of the measurement data obtained. In quantum mechanics,
however, measurement precision always comes at a price. In particular, 
infinite precision requires a measurement interaction which completely 
randomizes some of the unobserved system properties. Consequently, limited
precision may actually be desireable in quantum measurements.

For instance, a single mode of the electromagnetic field with a well defined 
photon number must
have a completely random phase. Therefore a precise photon number measurement
destroys phase coherence and all associated interference properties of
the field mode with other coherent modes. If phase coherence and
interference properties are preserved the intensity of the field mode can only 
be determined with a precision too low to resolve single photons. 
On the other hand, quantization emerges only when phase 
coherence is lost. Nevertheless a complete characterization of the 
light field dynamics requires both information about the intensity and 
the phase distributions. In general, it is therefore realistic to consider
a compromise between phase uncertainty and intensity uncertainty.

In quantum nondemoltion measurements of photon number, information about
the photon number $\hat{n}$ is obtained through the interaction of the 
measured field with a probe field \cite{Lev86,Fri92} or with probe atoms 
\cite{Bru90,Hol91}.
This interaction introduces phase noise into the measured system, as
required by the uncertainty relations \cite{Imo85,Kit87}. 
Since the purpose of the
procedure is a measurement of photon number, it is very tempting 
to assume that a perfect resolution of
photon number is the ideal case and therefore more desireable than 
a limited resolution. However, Kitagawa and coworkers \cite{Kit87}
have pointed out that even if photon number states are not resolved,
a quantum nondemolition measurement of photon number may produce
a minimum uncertainty state of phase and photon number. There is a 
trade off, then, between the noise introduced and the
resolution achieved, which requires the definition of a much larger 
class of ideal quantum measurements. By generalizing the conventional
projective measurement postulate, it is possible to investigate this
class of ideal quantum measurement, focussing especially on the transitional
regime between classical low noise measurements at low resolution
and the extreme quantum regime of fully resolved quantization and 
complete dephasing. 
It is shown in the following, that the statistical properties
of such intermediate resolution measurements include nonclassical 
correlations between the measured photon number and the phase noise introduced
in the measurement which can only be observed 
in this transitional regime.

In part \ref{sec:qnd}, a theoretical description of photon number measurements
with variable resolution is given and the effective measurement postulate is 
derived. In particular, the measurement operator provides a description of
the dephasing caused by the measurement interaction.

In part \ref{sec:emerge}, the statistics of the measurement results are
obtained. The transition from the classical limit to the quantum limit 
is discussed by pointing out the appearance of nonclassical correlations 
between the measurement result and the coherence after the measurement. 

In part \ref{sec:comp}, the correlations are compared to fundamental 
properties of the operator formalism. It is shown that the statistics
of the measurement results correspond to a specific operator ordering
in the evaluation of correlations.

In part \ref{sec:concl}, the results are summarized and possible implications
are discussed. It is argued that the measurement
statistics reveal that there is more to quantum reality than the integer 
photon number. By providing coherence, half integer photon numbers or
``fuzzy'' photon numbers also contribute to observable fact. 

\section{Variable resolution in ideal photon number measurements}
\label{sec:qnd}
\subsection{Light field quantization and measurement precision}
Based on the application of lasers, modern quantum optics has provided
a characterization of the quantum mechanical light field which is
much closer to a classical theory of noisy fields than the operator
formalism would suggest \cite{Mey99}. In particular, the classical
property of light field coherence is much easier to control than the 
nonclassical property of quantized photon number. It is indeed 
difficult to measure the exact photon number of a single, well defined 
light field mode. In multi-mode open systems such as lasers, Langevin
equations offer a better description of the light field dynamics than 
photon number rate equations, even in the presence of amplitude squeezing
\cite{Yam86}. This dominance of the classical wave properties in lasers
has motivated a new kind of criticism of the photon picture, expressed
especially in the notion of ``lasing without photons'' by Siegmann
\cite{Sie96,Sie95}. Even in the light of conventional quantum mechanics,
it is questionable whether the concept of photon number has any meaning 
before it is definitely measured. In particular, Heisenberg emphasized that 
no value can be assigned to a physical property if the system is not
in an eigenstate of that property \cite{Hei58}. After all, what photon number 
should be assigned to a coherent superposition of photon number states? 
It should be obvious that one cannot just pick out one eigenvalue while 
neglecting the others. Nevertheless, this point is so contrary to our 
natural intuition that it still raises controversies among physicists
\cite{PT99}. 

In a quantum nondemolition measurement of photon number, a nonlinear
coupling mechanism is utilized to shift a noisy and continuous pointer 
variable by an amount proportional to the photon number. As a consequence,
the measurement readout of the photon number measurement is generally 
both noisy and continuous. The discretess of the photon number
eigenvalues only emerges if the noise in the pointer variable is 
sufficiently low. Thus, the actual measurement result obtained is usually
a continuous variable and not a discrete one.
In order to study the emergence of photon number quantization, one should
therefore examine the properties of quantum measurements with variable
resolution 
and continuous values for the photon number measurement results. 
If the reality of integer photon numbers is somehow ``created''
in the measurement, there should be a transition from classical fields to
quantized fields depending only on the measurement resolution. 
While the basic tools for such an analysis are indeed provided by 
the standard quantum theory of measurement \cite{Neu55,Kit87}, the 
axiomatic nature of the mathematical approach often obscures the intuitive
classical limit. Therefore, it is useful to formulate a generalized
measurement postulate taking into account the limited measurement 
resolution. 
This measurement postulate summarizes the conventional results while
illustrating the fundamental aspects of coherence and decoherence more 
clearly, providing a shortcut to the derivation of quantum noise features.

\subsection{Generalized measurement postulate for pointer measurements}
In a quantum nondemolition measurement, a pointer variable $n_m$
of the probe system is shifted by an amount corresponding to the
photon number $n$ of the light field. However, since the
pointer variable $n_m$ is itself noisy, there is some error in this
procedure. Assuming Gaussian noise, the probability distribution
of $n_m$ subject to an uncertainty of $\delta\! n$ reads
\begin{equation}
\label{eq:probable}
P(n_m) = \left(2\pi\delta\! n^2\right)^{-1/2} 
         \exp\left(-\frac{(n-n_m)^2}{2\delta\! n^2}\right).
\end{equation}
This distribution applies to a photon number eigenstate. In order to
describe the effects of a measurement on superpositions of photon
number states, it is necessary to define an operater 
$\hat{P}_{\delta\! n} (n_m)$, such that the general effect of a measurement
result $n_m$ with a quantum mechanical uncertainty $\delta\! n$ on an
initial state $\mid \psi_i \rangle$ is given by $\hat{P}_{\delta\! n} (n_m)
\mid \psi_i \rangle$. The probability of obtaining the result $n_m$
and the state $\mid \psi_f (n_m) \rangle$ after the measurement
are then given by
\begin{eqnarray}
\label{eq:pworks}
P(n_m) &=& \langle \psi_i \mid \hat{P}^\dagger_{\delta\! n} (n_m)\;
         \hat{P}_{\delta\! n} (n_m)\mid \psi_i \rangle
\nonumber \\
\mid \psi_f (n_m) \rangle &=& \frac{1}{\sqrt{P(n_m)}} 
                \hat{P}_{\delta\! n} (n_m)
\mid \psi_i \rangle.
\end{eqnarray}

Note that the measurement thus described is ideal, since a pure state
remains pure and no additional decoherence is introduced.
It is assumed that the measurement system is prepared in a well defined
quantum state and that the readout is accurate. The source of the 
uncertainty in the measurement is the quantum noise in the 
pointer variable $n_m$ before the measurement interaction takes place.
By increasing this noise, the phase noise introduced in the measurement
interaction is reduced and vice versa. In a realistic situation, there
may be additional measurement uncertainties due to an inaccurate readout 
of the pointer or due to additional phase noise introduced in the measurement
interaction. Such additional noise sources cause decoherence and change
the pure state $\mid \psi_f (n_m) \rangle$ into a mixture which would have to
be represented by a density matrix. In the following, however, it is 
assumed that such additional noise sources can be avoided.
It is then possible to deduce the correct measurement operator by 
comparing equations (\ref{eq:probable}) and (\ref{eq:pworks}).
It reads
\begin{equation}
\label{eq:postulate}
\hat{P}_{\delta\! n}(n_m) = \left(2\pi\delta\! n^2\right)^{-1/4} 
         \exp\left(-\frac{(\hat{n}-n_m)^2}{4\delta\! n^2}\right).
\end{equation}
This operator describes the relation of the photon number operator $\hat{n}$ 
with the value $n_m$ obtained in the measurement. Thus the connection between
the quantum system and the classical measurement readout is established.
Although the standard measurement postulate as formulated by von Neumann
\cite{Neu55} can be recovered by either letting $\delta\! n$ approach zero
or by applying $\hat{P}_{\delta\! n} (n_m)$ many times, the generalized concept
of measurement represented by $\hat{P}_{\delta\! n} (n_m)$ describes a much 
wider range of physical situations and is definitely closer to the 
kind of perception we know from everyday experience. In particular, 
it describes the classical limit of the uncertainty relations
in the case of low resolution, $\delta\! n \gg 1$.

\subsection{Photon number squeezing and phase noise}
Although, strictly speaking, the phase of a light field mode is not an
observable since no phase operator can be constructed, approximate
operators and phase space distributions show that there is an uncertainty
relation between photon number and phase given by $\delta n \delta \phi \geq
1/2$ \cite{Sus64,Peg89}. The role of this uncertainty in quantum nondemolition 
measurements of photon number has been investigated in the context of 
measurements using the optical Kerr effect \cite{Imo85,Kit87}. 
It will be shown in the following that the generalized 
measurement operator $\hat{P}_{\delta\! n}(n_m)$ faithfully reproduces these 
experimentally confirmed results. 

Since the phase itself cannot be
represented by an operator, it is more realistic to illustrate the 
decoherence induced by the phase noise by analyzing the reduction in 
the expectation value of the complex
field amplitude $\langle \hat{a} \rangle$. Adding Gaussian phase noise
with a variance of $\delta\!\phi^2$
to an arbitrary field state reduces the initial 
expectation value of the 
amplitude $\langle \hat{a} \rangle_i$ to a final value of 
\begin{equation}
\label{eq:phaseconvert}
\langle \hat{a} \rangle_f = \exp\left(-\frac{\delta\!\phi^2}{2}\right)
\langle \hat{a} \rangle_i.
\end{equation}
The overall average 
$\langle \hat{a} \rangle_f (\mbox{av.})$ of the field expectation 
value after the measurement is given by
\begin{eqnarray}
\label{eq:decoher}
\langle \hat{a} \rangle_f (\mbox{av.}) &=&
\int \langle \psi_f (n_m) \mid \hat{a} \mid \psi_f (n_m) \rangle
P(n_m) dn_m
\nonumber \\ &=&
\int 
\langle \psi_i \mid \hat{P}_{\delta\! n} (n_m) \hat{a}
                    \hat{P}_{\delta\! n} (n_m)\mid \psi_i \rangle
dn_m
\nonumber \\ &=&
\exp\left(- \frac{1}{8\delta\! n^2}\right) 
    \langle \psi_i \mid \hat{a} \mid \psi_i \rangle
.
\end{eqnarray}
According to equation (\ref{eq:phaseconvert}), this reduction in 
amplitude corresponds to a Gaussian phase noise with a variance of
\begin{equation}
\label{eq:phasenoise}
\delta \! \phi^2 = \frac{1}{4\delta\! n^2}.
\end{equation}
Thus the amount of phase noise introduced in the measurement
corresponds to the minimum noise required by the uncertainty relation
of phase and photon number for a measurement resolution of $\delta\! n$.
This is a direct consequence of assuming an ideal quantum mechanical 
measurement which does not introduce additional phase noise. In a 
realistic situation, it is likely that the phase noise introduced is 
somewhat higher than this ideal quantum limit. Relation (\ref{eq:phasenoise})
may then be used to determine how much excess phase noise is introduced 
in a given experimental setup. 
Note that this excess noise may originate not only from an additional 
source of decoherence, but also from an inaccurate readout of the pointer 
variable. 

\section{The emergence of quantization}
\label{sec:emerge}
\subsection{Measurement of a coherent state}
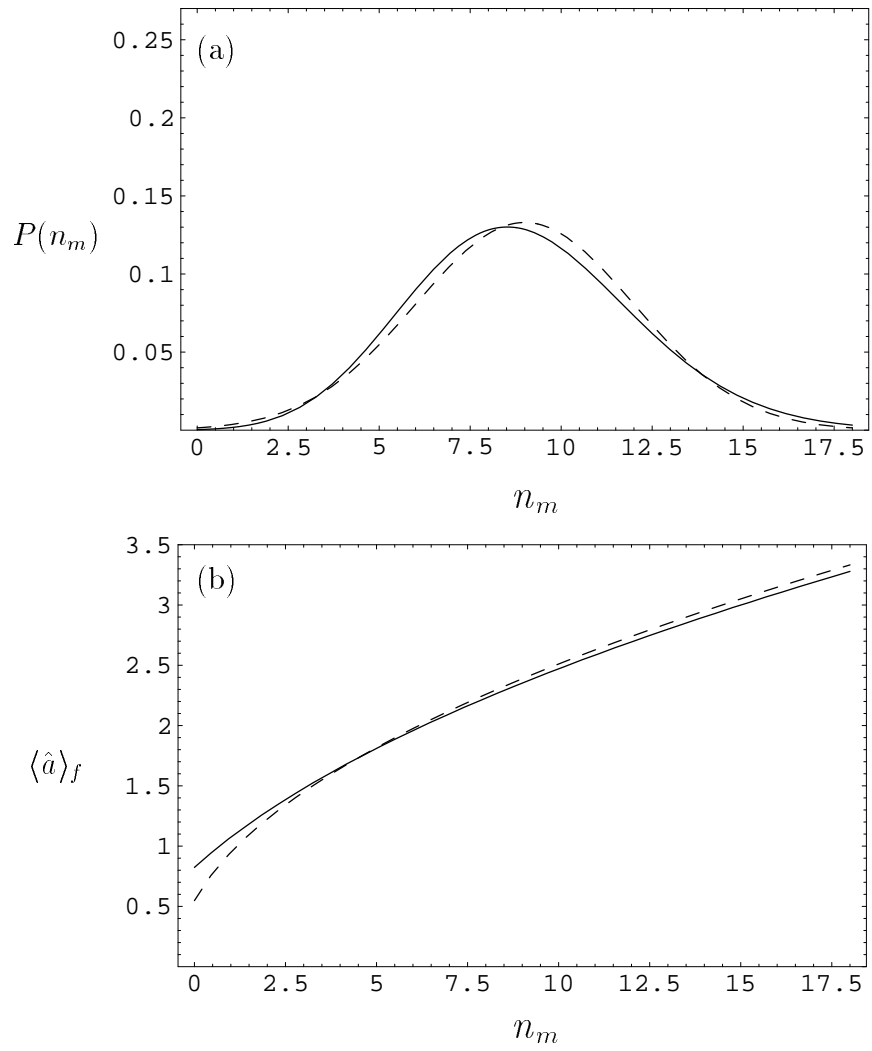
\begin{figure}
\caption{\label{classic}
Photon number measurement statistics of a coherent state with 
an average amplitude of $\alpha=3$ at a photon number resolution of
$\delta\! n=0.7$. (a) shows the probability distribution over measurement
results $n_m$. Quantization is not resolved yet. The dashed curve corresponds
to the approximate result using a Gaussian photon number distribution as
explained in the text. (b) shows the expectation value 
$\langle\hat{a}\rangle_f$ after the measurement as a function of the 
measurement result $n_m$. The dashed curve is the result obtained by
multiplying a coherent amplitude of $\sqrt{n_m+1/2}$ with the dephasing 
factor.}
\end{figure}

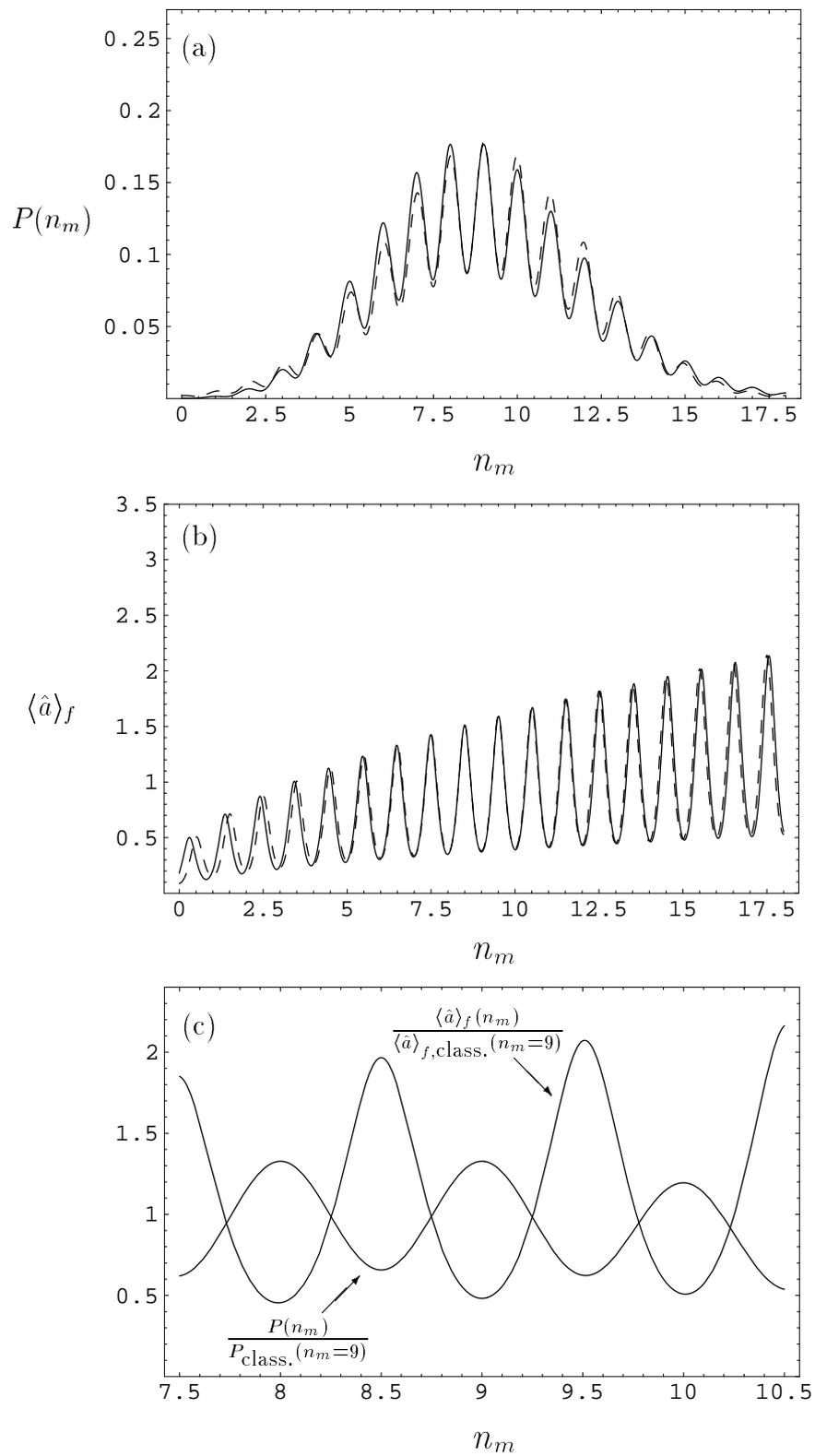
\begin{figure}
\caption{\label{highmod}
Photon number measurement statistics of a coherent state with 
an average amplitude of $\alpha=3$ at a photon number resolution of
$\delta\! n=0.3$. (a) to (c) are as in the previous figure.}
\end{figure}

If the initial state $\mid \psi_i \rangle$ is a coherent state
$\mid \alpha \rangle$ with the photon number state expansion
\begin{equation}
\mid \alpha \rangle = \exp(-\frac{|\alpha|^2}{2}) \sum_n 
        \frac{\alpha^n}{\sqrt{n!}} \mid n \rangle,
\end{equation}
then the measurement statistics 
defined by equation (\ref{eq:pworks}) reads
\begin{eqnarray}
\label{eq:xPresult}
P(n_m) &=& \langle \alpha \mid \hat{P}_{\delta\! n}^2(n_m) \mid \alpha \rangle
\nonumber \\[0.3cm] &=& \frac{\exp(-|\alpha|^2)}{\sqrt{2\pi\delta\! n^2}}
      \sum_n \frac{|\alpha|^{2n}}{n!} 
             \exp\left(-\frac{(n-n_m)^2}{2\delta\! n^2}\right),
\end{eqnarray}
and the coherent amplitude
$\langle \hat{a} \rangle_f$ after the measurement reads
\begin{eqnarray}
\label{eq:xCresult}
\langle \hat{a} \rangle_f(n_m) &=& 
\frac{\langle \alpha \mid \hat{P}_{\delta\! n}(n_m)
             \hat{a} \hat{P}_{\delta\! n}(n_m) \mid \alpha \rangle}
     {\langle \alpha \mid \hat{P}_{\delta\! n}^2(n_m) \mid \alpha \rangle}
\nonumber \\
&=& \alpha \; \exp\left(-\frac{1}{8\delta\! n^2}\right)
\frac{\sum_n \frac{|\alpha|^{2n}}{n!} 
             \exp\left(-\frac{(n+\frac{1}{2}-n_m)^2}{2\delta\! n^2}\right)}
     {\sum_n \frac{|\alpha|^{2n}}{n!} 
             \exp\left(-\frac{(n-n_m)^2}{2\delta\! n^2}\right)}.
\end{eqnarray} 
The results shown in figures \ref{classic} to \ref{quantumlimit} have
been calculated using these exact results. However, it is helpful to apply
some approximations in order to identify the quantization effects. 

For $|\alpha|^2\gg 1$, the photon number distribution may be approximated
by a Gaussian distribution with a mean photon number $|\alpha|^2$ and a 
photon number fluctuation of $|\alpha|$. The application of the measurement 
operator $\hat{P}_{\delta\! n}(n_m)$ then results in a convolution of 
two Gaussians. If the resolved photon number $\delta\! n$ is much smaller 
than the photon number fluctuation $|\alpha|$, then the amplitude of the
photon number state components of $\mid \alpha\rangle$ does not change much 
within the measurement interval of $n_m\pm \delta\! n$ and the convolution 
may be approximately factorized into a product reading 
\begin{eqnarray}
\label{eq:Pstate}
\hat{P}_{\delta\! n}(n_m) \mid \alpha \rangle &\approx& 
\overbrace{(2\pi |\alpha|^2)^{-1/4} 
\exp\left(-\frac{(n_m-|\alpha|^2)^2}{4|\alpha|^2}\right)}^{
         \mbox{Gaussian intensity distribution of $\mid \alpha \rangle$}}
\nonumber \\
&& \times \underbrace{\sum_n (2\pi \delta\! n^2)^{-1/4} 
\exp\left(-\frac{(n-n_m)^2}{4\delta\! n^2}\right) 
\exp\left(-i\phi n\right) \mid n \rangle}_{\mbox{decoherence 
and quantization effects}},
\end{eqnarray}
where the phase $\phi$ is defined by $\alpha = |\alpha|\exp(-i\phi)$.

It is thus possible to separate the state dependent photon number distribution
from the fundamental effects of decoherence and quantization. By applying the
approximations of equation (\ref{eq:Pstate}) to the measurement statistics
described by equations (\ref{eq:xPresult}) and (\ref{eq:xCresult}), 
an even clearer separation of
classical noise properties and quantization effects is obtained. The 
approximate results read
\begin{equation}
\label{eq:nmprob}
P(n_m) 
\approx
\underbrace{(2\pi |\alpha|^2)^{-1/2} 
\exp\left(-\frac{(n_m-|\alpha|^2)^2}{2|\alpha|^2}\right)}_{\mbox{
                                classical intensity distribution}}\; 
\underbrace{\sum_n (2\pi \delta\! n^2)^{-1/2} 
\exp\left(-\frac{(n-n_m)^2}{2\delta\! n^2}\right)}_{\mbox{
                                           quantization effects}}
\end{equation}
for the probability, and
\begin{equation}
\label{eq:nmdeco}
\langle \hat{a}\rangle_f (n_m) \approx 
\underbrace{\exp\left(-i\phi\right) \sqrt{n_m+\frac{1}{2}} \;
\exp\left(-\frac{1}{8\delta\! n^2}\right)}_{\mbox{
                               classical amplitude average}} \;
\underbrace{\frac{\sum_n 
     \exp\left(-\frac{(n-\frac{1}{2}-n_m)^2}{2\delta\! n^2}\right)}
         {\sum_n \exp\left(-\frac{(n-n_m)^2}{2\delta\! n^2}\right)}}_{\mbox{
                                                       quantization effects}}
\end{equation}
for the coherent amplitude. Note that only the phase of the coherent
amplitude expectation value $\langle \hat{a}\rangle_f$ after the 
measurement depends on the initial value of $\alpha$. The absolute value
is determined by the measurement result and is proportional to 
$\sqrt{n_m+1/2}$. This result corresponds to the classical notion that 
the absolute value of the coherent amplitude should be the square root of the
intensity. 

The sums which express the quantization effects in equations 
(\ref{eq:nmprob}) and (\ref{eq:nmdeco}) are periodic functions of
$n_m$. In other words, quantization effects only depend on how close the
measurement result $n_m$ is to an integer value. Because of this periodicity, 
the sums can be expressed as Fourier series. Specifically,
\begin{equation}
 (2\pi \delta\! n^2)^{-1/2}\sum_n 
\exp\left(-\frac{(n-n_m)^2}{2\delta\! n^2}\right)
= 1 + 2 \sum_{k=1}^{\infty} \exp\left(-2\pi^2\delta\! n^2 k^2\right)
 \cos \left(2\pi k n_m\right)
\end{equation}
and
\begin{equation}
(2\pi \delta\! n^2)^{-1/2}\sum_n 
\exp\left(-\frac{(n-\frac{1}{2}-n_m)^2}{2\delta\! n^2}\right)
= 1 - 2 \sum_{k=1}^{\infty} \exp\left(-2\pi^2\delta\! n^2 k^2\right)
 \cos \left(2\pi k n_m\right)
.
\end{equation}
Note that the Fourier coefficients are Gaussians in the modulation
frequency variable $k$. The high frequency components of the periodic
modulations are therefore strongly
suppressed. Depending on the measurement resolution $\delta\! n$, it
is reasonable to limit the expansion to only the first few 
contributions. This resolution dependent truncation of the Fourier
series defines the transition from the classical regime to the 
quantum regime.

\subsection{From the classical limit to full quantization}
In the classical limit, all Fourier components with $k>1$ are
negligible.
The measurement probability and the 
expectation value of the coherent field after the measurement read
\begin{eqnarray}
\label{eq:classlimit}
P_{\mbox{class.}}(n_m) &=& 
(2\pi |\alpha|^2)^{-1/2} 
\exp\left(-\frac{(n_m-|\alpha|^2)^2}{2|\alpha|^2}\right)
\nonumber \\
\langle \hat{a}\rangle_{f,\mbox{class.}} (n_m) &=& 
\sqrt{n_m+1/2} \; 
\exp\left(-i\phi\right)\exp\left(-\frac{1}{8\delta\! n^2}\right)
.
\end{eqnarray}
These results correspond to the classical assumption of continuous 
light field intensity and equally continuous Gaussian noise in the 
light field phase and amplitude. 
A typical example is shown in figure \ref{classic} for a coherent
state with an amplitude of $\alpha=3$. 
The measurement resolution is at $\delta\! n=0.7$, quite close to the
quantum limit. Nevertheless, the approximate results of equations 
(\ref{eq:classlimit}) correspond quite well to the more precise results
of equations (\ref{eq:xPresult}) and (\ref{eq:xCresult}). Indeed, the 
main discrepancy between the probability 
distribution $P(n_m)$ given by equation (\ref{eq:classlimit})
and the exact result is due to the asymmetry of the Poissonian photon
number distribution which has been neglected by assuming a Gaussian
photon number distribution in equations (\ref{eq:nmprob}) and 
(\ref{eq:nmdeco}). This deviation gets much smaller as the 
average photon number of the coherent state is increased. However,
it is already a good approximation at the average photon number 
of nine shown in the examples.

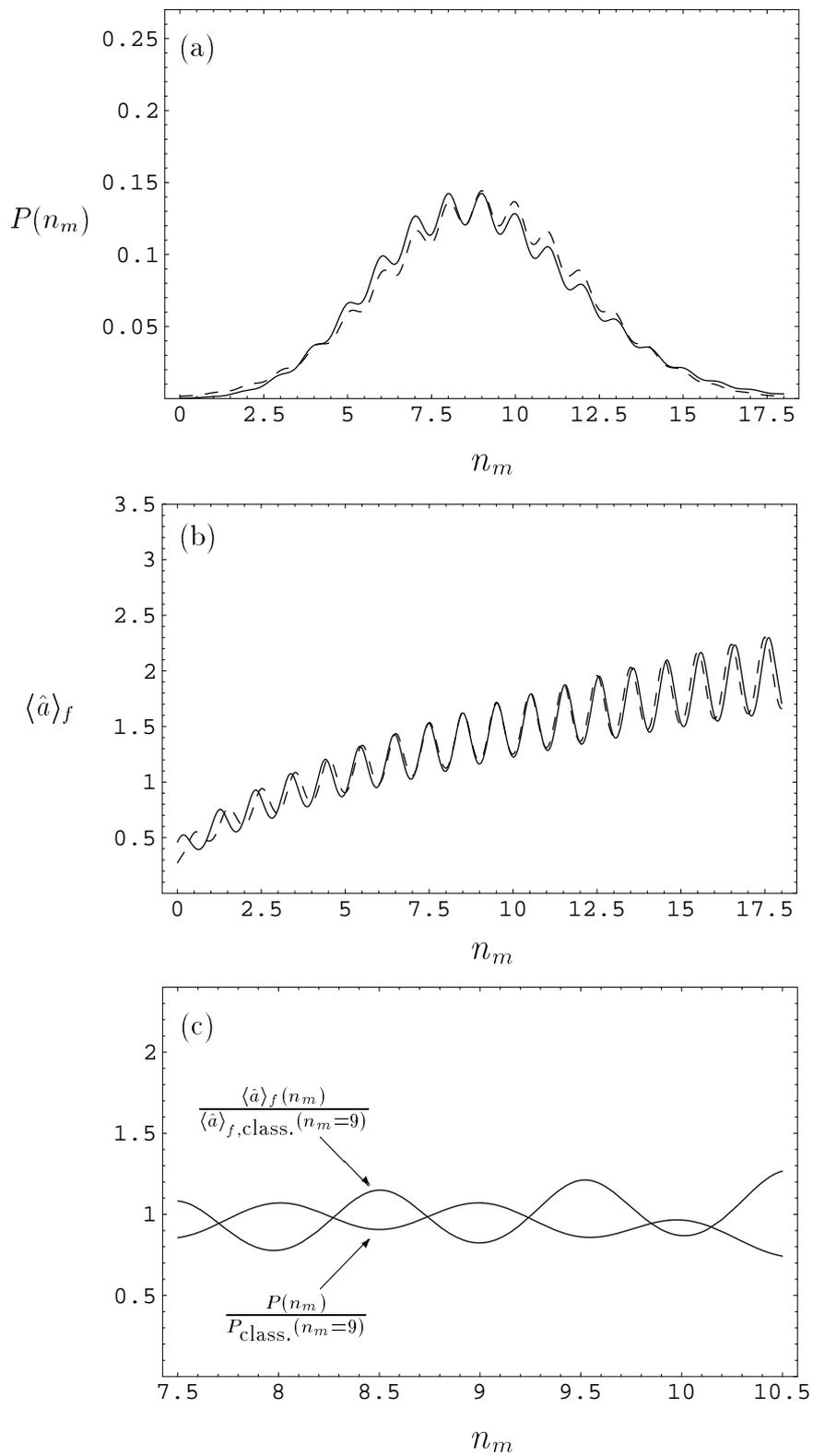
\begin{figure}
\caption{\label{lowmod} Photon number measurement statistics of a 
coherent state with 
an average amplitude of $\alpha=3$ at a photon number resolution of
$\delta\! n=0.4$. (a) shows the probability distribution over measurement
results and (b) shows the expectation value $\langle\hat{a}\rangle_f(n_m)$ 
after the measurement. The dashed curves correspond to the approximate
formulas given in the text. (c) shows details of the quantum mechanical
modulations of measurement probability and coherence after the measurement
near $n_m=9$, normalized by the respective classical results
$P_{\mbox{class.}}(n_m=9)$ and 
$\langle\hat{a}\rangle_{f,\mbox{class.}}(n_m=9)$.}
\end{figure}

As the quantum limit is approached, the classical results are modulated
by quantum effects. In the probability distribution of measurement
results, this modulation appears as a fringe pattern similar to that
caused by an interference effect. At the same time, a complementary fringe
pattern emerges in the coherence after the measurement as given by
$\langle \hat{a}\rangle_f (n_m)$. The lowest order contributions to these
quantization effects read
\begin{eqnarray}
\label{eq:loworder}
P(n_m) &=& P_{\mbox{class.}}(n_m)
\left(1+2 \exp\left(-2\pi^2\delta\! n^2 \right)
 \cos \left(2\pi n_m\right)\right)
\nonumber \\
\langle\hat{a}\rangle_f (n_m) &=& 
      \langle\hat{a}\rangle_{f,\mbox{class.}} (n_m)
\frac{1-2 \exp\left(-2\pi^2\delta\! n^2\right)
 \cos \left(2\pi n_m\right)}{1+2 \exp\left(-2\pi^2\delta\! n^2\right)
 \cos \left(2\pi n_m\right)}
.
\end{eqnarray}
The accuracy of this approximation is worst for 
$\langle\hat{a}\rangle_f (n_m)$ at integer or half integer values of 
$n_m$. At these points, it is accurate to within 1\% for 
$\delta\! n \geq 0.27$ and accurate to within 10\% for $\delta\! n \geq 0.23$.
Thus, the reliability of the lowest order approximation is generally very 
high above $\delta\! n \approx 0.25$.
Figure \ref{lowmod} shows the probability 
distribution and the coherent amplitude after the measurement at a resolution
of $\delta \!n = 0.4$. This resolution corresponds to a modulation factor 
of $2 \exp(-2\pi^2 \delta\! n^2)= 0.085$. The modulation is still very weak
and the likelihood of obtaining an integer result is only about 1.2 times 
higher than the likelihood of obtaining a half integer result. Nevertheless, 
the quantization fringes in $P(n_m)$ and the decoherence fringes in 
$\langle\hat{a}\rangle_f (n_m)$ are clearly visible. The anticorrelation
of the probability peaks and the coherence maxima is illustated in 
figure \ref{lowmod} c) which shows the respective modulations near $n_m=9$, 
normalized using the classical results at $n_m=9$. 
Figure \ref{highmod} shows the probability 
distribution and the coherent amplitude after the measurement at a resolution
of $\delta \!n = 0.3$. This resolution corresponds to a modulation factor 
of $2 \exp(-2\pi^2 \delta\! n^2)= 0.338$. 
The likelihood of obtaining an integer result is about twice as high
as that of obtaining a half integer result and the reduction in the coherent
amplitude is about four times greater for integer $n_m$ than for half integer
$n_m$. At an average decoherence factor of $\exp(-1/(8\delta\! n^2))=0.25$,
the average coherent amplitude after the measurement is still quite 
significant. A measurement resolution of $\delta\! n=0.3$ thus combines
aspects of photon number quantization and aspects of phase coherence,
defining the center of the transitional regime between continuous 
field intensities and quantized photon numbers.

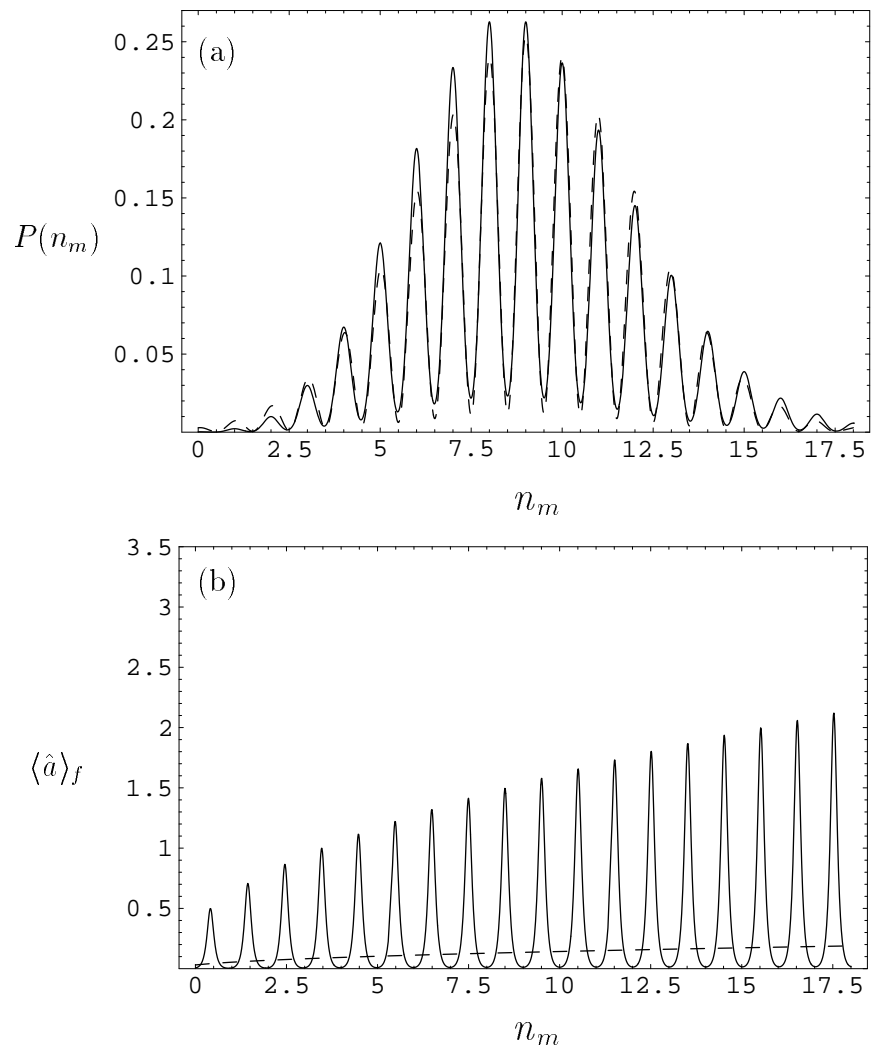
\begin{figure}
\caption{\label{quantumlimit} Photon number measurement statistics of a 
coherent state with 
an average amplitude of $\alpha=3$ at a photon number resolution of
$\delta\! n=0.2$. (a) shows the probability distribution over measurement
results. The dashed curve correspond to the approximate
formulas given in the text. (b) shows the expectation value 
$\langle\hat{a}\rangle_f(n_m)$ after the measurement. The dashed curve 
shows the classical result, $\langle\hat{a}\rangle_{f,\mbox{class.}}(n_m)$.}
\end{figure}

Between a resolution of $\delta\! n=0.3$ and a resoltuion of 
$\delta\! n=0.2$, the approximation given by equation (\ref{eq:loworder})
breaks down. For $\delta\! n<0.2$, the probability distribution is
given by isolated Gaussians centered around integer measurement results
$n_m$. Half integer results become extremely unlikely. However, if such
an unlikely result is obtained, there still is coherence even in extremely
precise measurments. This fact is usually obscured by the assumption of
infinite precision inherent in the conventional projective measurement 
postulate. Figure \ref{quantumlimit} shows the probability distribution
and the coherence after the measurement for a resolution of
$\delta \! n=0.2$. Note that the approximation given by 
equation (\ref{eq:loworder}) is still very 
good for the probability distribution. However, the relative error in the
peak values of the coherent amplitude $\langle\hat{a}\rangle_f$ after the
measurement is nearly 100\%. Therefore, the dashed curve in figure
\ref{quantumlimit} b) does not show the approximate result, but instead
shows the classical approximation $\langle \hat{a}\rangle_{f,\mbox{class.}}$ 
given by equation (\ref{eq:classlimit}). This comparison illustrates the
relatively high coherence at half integer measurement results $n_m$. 
At half integer measurement results $n_m$, the 
expectation value $\langle \hat{a}\rangle_{f,\mbox{class.}}$ of the coherent
amplitude is equal to $(\sqrt{n_m+1/2})/2$, or one half of the amplitude
corresponding to a classical light field intensity of $n_m+1/2$.
This result is valid for all $\delta\!n<0.2$, regardless of the average
dephasing induced by the measurement interaction. Therefore, the peak
values of the coherence after the measurement are much higher than the
classical results, while the minima at integer photon number are actually
closer to zero than the classical interpretation of dephasing
would suggest. In the case of $\delta \! n=0.2$ shown in figure 
\ref{quantumlimit}, the classical approximation predicts an average 
decoherence factor of $\exp(-1/(8\delta\! n^2))=0.044$. However, the
peak values of coherence at half integer photon number are more than 
ten times higher and the minima at integer photon number are more than 
ten times lower than the classically expected coherence after dephasing.
Since the likelihood of integer results is about ten times higher than the 
likelihood of half integer results, the main contribution to the average 
coherence after the measurement still originates from half integer photon 
number results. Even at fully resolved quantization, the half integer
phopton number esults thus provide a contribution to the dephasing
statistics.

\subsection{Correlation between quantization and dephasing}
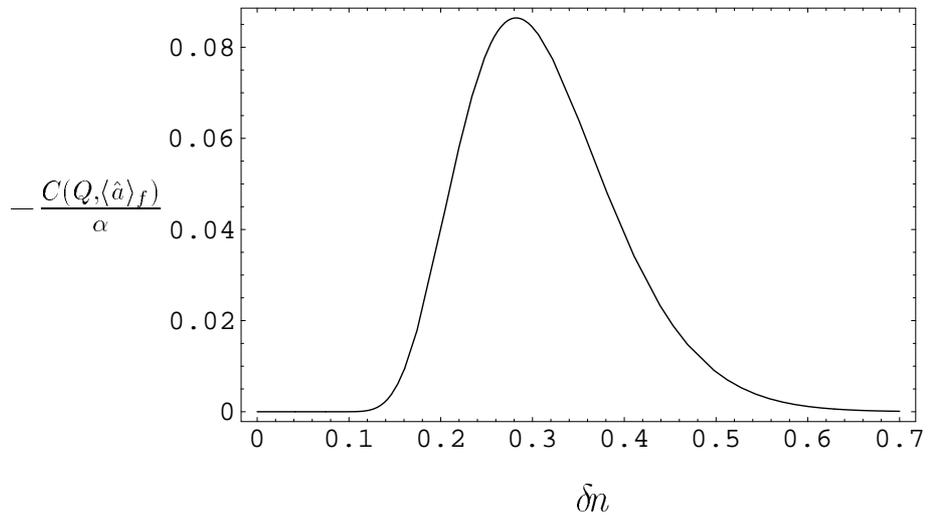
\begin{figure}
\caption{\label{corrfig}Normalized anticorrelation of the quantization 
$Q$ of the measurement result
$n_m$ and the coherence $\langle \hat{a} \rangle_f(n_m)$ after the measurement
as a function of measurement resolution $\delta \! n$.}
\end{figure}

The discussion above reveals a clear qualitative difference between
measurement results $n_m$ of integer photon number and of half integer
photon number. To obtain a quantitative expression, it is necessary to
define a measure of quantization associated with each measurement
result $n_m$. In the following, the quantization $Q$ of a measurement 
result $n_m$ is therefore defined as 
\begin{equation}
Q (n_m) = \cos\left(2\pi n_m\right).
\end{equation}
Thus, the quantization $Q$ of integer values of $n_m$ is +1 and the
quantization of half integer values is -1. In the classical case, this
results in an average quantization of zero. The average quantization
$\bar{Q}$ of the measurement results is given by
\begin{eqnarray}
\bar{Q} &=& \int dn_m \; Q(n_m) P(n_m)
\nonumber \\
        &=& \exp\left(-2\pi^2\delta\! n^2\right).
\end{eqnarray}
Since $\bar{Q}$ depends only on $\delta\! n$, it may be used as an
experimental measure of the resolution obtained in quantum nondemolition
measurements of photon number.
It is now possible to evaluate the correlation between the quantization
observed and the coherence after the measurement by averaging the
product,
\begin{eqnarray}
\overline{Q \langle \hat{a}\rangle_f} &=& 
\int dn_m \; Q(n_m)\langle \hat{a}\rangle_f(n_m) P(n_m)
\nonumber \\
        &=& - \exp\left(-2\pi^2\delta\! n^2\right) 
              \exp\left(-\frac{1}{8 \delta\! n^2}\right) \alpha
\nonumber \\
        &=& - \bar{Q} \;  \langle\hat{a}\rangle_f(\mbox{av.}).
\end{eqnarray}
The average of the product of quantization and coherence is exactly equal 
to the negative product of the averages. Therefore, quantization and
coherence are strongly anti-correlated. The correlation 
$C(Q,\langle \hat{a}\rangle_f)$ is given by
\begin{eqnarray}
\label{eq:corr}
C(Q,\langle \hat{a}\rangle_f) &=&
  \overline{Q \; \langle \hat{a}\rangle_f} 
- \bar{Q} \; \langle \hat{a}\rangle_f (\mbox{av.}) 
\nonumber \\[.3cm]
&=& -2 \bar{Q} \; \langle \hat{a}\rangle_f (\mbox{av.})
\nonumber \\
&=& - 2 \exp\left(-2\pi^2\delta\! n^2\right) 
        \exp\left(-\frac{1}{8 \delta\! n^2}\right) \alpha.
\end{eqnarray}
Figure \ref{corrfig} shows this correlation as a function of measurement 
resolution $\delta \! n$. 
The correlation is maximal at 
$\delta\! n= 1/(2\sqrt{\pi})$, which is a resolution of about 
0.282 photons. At this point, the average quantization $\bar{Q}$
is equal to $\exp(-\pi/2)=0.208$ and the average coherent amplitude
$\langle \hat{a}\rangle_f(\mbox{av.})$ after the measurement is 
equal to $\exp(-\pi/2)=0.208$ times the original amplitude $\alpha$.

There appears to be a well defined transition from the 
classical limit to the quantum limit of measurement resolution 
at $\delta\! n = 1/(2\sqrt{\pi})$, which is characterized by statistical
properties not observable in either the extreme quantum limit or in the
classical limit. Since it should be possible to obtain these statistical
properties from experimental results, some measure of reality must be 
attributed to the concept of variable quantization $Q$. Specifically, 
even though it is clear that only measurement results
of full quantization $Q=1$ remain as the resolution is increased, 
the reduced decoherence at $Q=-1$ demonstrates that such results can not
be interpreted as measurement errors due to either a higher or a lower
photon number. This measurement scenario thus highlights the problem of
assuming the existance of an integer photon number before the photon 
number is actually measured. Obviously, quantization is not a property 
of the system which is simply hidden by the noise of the low precision 
measurement in the classical limit. Some very real physical properties are
associated with noninteger values of photon number measurement results.
Possibly, it is necessary to consider operator values other than the
eigenvalues as part of the physical reality associated with quantum
mechanical operator variables.

\section{Fundamental properties of the operator formalism}
\label{sec:comp}
\subsection{Quantization and the parity operator}
\label{sec:parity}
The generalized measurement operator $\hat{P}_{\delta\! n}(n_m)$
describes both classical and quantum mechanical features of measurements
in terms of a quantum mechanical operator. Classically, it would be possible 
to distinguish between the measurement result $n_m$ and the actual photon
number $n$. In quantum mechanics, however, the photon number $\hat{n}$ 
is an operator which does not have a well defined value unless the field 
is in a photon number eigenstate. Therefore, the relationship between
the measurement result $n_m$ and the photon number operator $\hat{n}$
is quite different from the classical relationship between a noisy
measurement result and the true value of the measured quantity.

A quantum mechanical property which may provide a connection between
the definition of quantization $Q$ based on the measurement result $n_m$ 
and the properties of the photon number operator $\hat{n}$ is the
parity $\hat{\Pi}$ defined as
\begin{equation}
\hat{\Pi} = (-1)^{\hat{n}}.
\end{equation}
The square of the parity $\hat{\Pi}^2$ may then be associated with the
quantization $Q$.
Of course, the quantum mechanical value of quantization is always one. 
However, by ``breaking apart'' the square of the parity, a correlation
between quantization and coherent field amplitude may be established.
It reads
\begin{equation}
\label{eq:pcorr}
\langle \hat{\Pi} \hat{a} \hat{\Pi} \rangle 
- \langle \hat{\Pi}^2 \rangle \langle \hat{a} \rangle 
= - 2 \langle \hat{\Pi}^2 \rangle \langle \hat{a} \rangle.
\end{equation}
If $\langle \hat{\Pi^2} \rangle$ is identified with $\bar{Q}$ and 
$\langle \hat{a} \rangle$ is identified with 
$\langle \hat{a} \rangle_f(\mbox{av.})$,
this correlation corresponds to the one given in equation (\ref{eq:corr}).
The relationship between coherence and quantization can thus be traced
to the anti-commutation between parity and field amplitude, 
$\hat{\Pi}\hat{n}=-\hat{n}\hat{\Pi}$. One could indeed argue that the
correlation which appears in the measurement is hidden in the commutation
relations of the operator formalism.

\subsection{Ambiguous correlations in the operator formalism}
The correlation given in equation (\ref{eq:pcorr}) is of course a result
of the specific order in which the operators have been applied. Since
$\hat{\Pi}^2$ is always one, there is no correlation as soon as both
parity operators are placed on the same side of the field operator 
$\hat{a}$. In principle, it is not possible to determine the correlation
between noncommuting quantum variables directly from the operator 
formalism because of this ambiguity concerning the ordering of the
operators. 

In particular, the case of photon number quantization and parity 
belongs to a general class of correlations based on the inequality
\begin{equation}
\label{eq:ineq}
\frac{1}{2}\langle \hat{A}\hat{B}^2 + \hat{B}^2\hat{A}\rangle
\neq \langle \hat{B}\hat{A}\hat{B} \rangle,
\end{equation}
where $\hat{A}$ and $\hat{B}$ represent arbitrary noncommuting operator
variables. The operator ordering $\hat{B}\hat{A}\hat{B}$ allows correlations
even if the quantum state is an eigenstate of $\hat{A}$ or $\hat{B}^2$.
This property definitely contradicts any assumption of classical statistics.
Nevertheless, such correlations can be obtained in experiment, even though
the outcome of a direct measurement of $\hat{A}$ or $\hat{B}$ performed on 
the initial state would be perfectly predictable. 
Thus the quantum nondemolition measurement discussed in this paper
represents an example of a more general class of measurements revealing
fundamental nonclassical properties of quantum statistics.

\subsection{Operator ordering and physical reality}
In the theory of quantum mechanics, the classical values of physical
variables are replaced by operators. Consequently, it is not possible to
assign a well defined value to an operator variable if the system is not in
an eigenstate of the operator. This situation calls for a review of our
concepts of physical reality, as can be seen from the arguments concerning
entanglement and the debate of hidden variables \cite{PT99,PT90}. 
Quantum mechanical uncertainty is definitely quite different from
a classical lack of knowledge, and this difference is revealed in the
correlations between noncommuting variables. For instance, the EPR 
argument basically uses the entanglement of two particles to establish 
a correlation between position and momentum of the same particle - thus 
trying to circumvent the restrictions imposed by uncertainty on 
Einsteins arguments in the Bohr-Einstein dialogue \cite{BE35}.
However, as Bell has shown, the correlations between noncommuting
variables thus obtained cannot be represented by a classical probability
distribution \cite{Bel64}. Since this paradox is an inherent property of
the operator formalism, it should be possible to trace its origin directly
to the fundamental nonclassical properties of quantum mechanical measurements.

In principle it would be desireable to know the value of a correlation 
between noncommuting variables such as the parity $\hat{\Pi}$ and the coherent
amplitude $\hat{a}$ without reference to a measurement. If there were 
hidden variables defining classical values for both operators, there
should also be a well defined correlation. However, the formalism itself
introduces an ambiguity. A formal calculation of correlations based
on the expectation values of operator products raises the question of 
operator ordering. A particularily striking ambiguity is represented by
equation (\ref{eq:pcorr}), since it permits a correlation of $\hat{\Pi}^2$
with the coherent amplitude even though the eigenvalues of $\hat{\Pi}^2$
are all one. Of course one could argue that it should not be allowed to
separate the square of the parity operator. However, such a postulate
would not be based on any physical observation but only on preconceived
notions of what reality should be like. It is therefore important
to note that unusual correlations such as the one given by equation
(\ref{eq:pcorr}) can have a real physical meaning in measurement statistics.

Since quantum mechanics does not allow the simultaneous assignment of 
well defined physical values to noncommuting observables, it is not possible
to discuss correlations between such observables without a definition
of the measurement by which such correlations are obtained. The futility of 
trying a more general approach is clearly revealed by the ambiguity of 
the correlations caused by the commutation relations between operators.

\section{Conclusions and Outlook}
\label{sec:concl}

\subsection{Interpretation of the nonclassical correlations}
The results presented above show that a quantum
nondemolition measurement reveals much more than just the photon number
of a light field at an intermediate measurement resolution close
to $\delta\! n=0.3$. In this intermediate regime, the property that phase
coherence in the field requires quantum coherence  
between neighbouring photon number states emerges visibly as a 
correlation between the continuous measurement result $n_m$
and the coherence after the measurement $\langle \hat{a} \rangle_f$.
This measurement scenario thus reveals the difference between
quantum mechanical uncertainty and a classical lack of precision.
In particular, there is a real physical difference between the
measurement results of half integer photon number and the measurement
results of integer photon number which makes it impossible to 
argue that the measurement of half integer photon number is merely
an error. By introducing the variable $Q$ to denote the quantization
of the measurement result, it is possible to evaluate the correlation
between quantization and decoherence in the measurement. 
In the operator formalism, the quantization can be interpreted
as the square of the parity operator $\hat{\Pi}$. It is then possible 
to derive the observed correlation directly from the operator formalism.

The correlation obtained both from the statistics of the quantum 
nondemolition measurement and from the operator statistics
suggests the reality of half integer photon number results.
Depending on the circumstances, quantum measurements may therefore 
reveal physical values of operator variables which are quite 
different from the eigenvalues of the corresponding operators. 
At the same time, the ambiguity of the correlations between operator
variables shows that an identification of neither eigenvalues $n$
nor measurement results $n_m$ with elements of reality can be valid.
It is therefore not suficient to extend the range of photon number
values. Instead, the statistics of physical properties should be
based on the measurement results obtained in a specific measurement
setup. The ambiguity in the formalism can then be resolved by applying
the appropriate generalized measurement postulate.

It seems that the physical property of light field intensity given
by the photon number can not be attributed to any measurement independent
elements of reality.
Possibly, it might be a useful compromise to regard the measurement results
$n_m$ as elements of a fundamentally noisy reality, while acknowledging the
qualitative dependence of the measurement result on the resolution 
$\delta\! n$. In the classical limit, the identification of $n_m$ with
the actual light field intensity is usually not problematic. Therefore, 
our classical concept of reality survives on the macroscopic level,
even though it has to be abondoned in the microscopic regime.
In the quantum limit, $n_m$ can again be identified with the eigenvalues
of the operator $\hat{n}$. In this manner, a continuous transition between 
our classical concept of reality and the mysterious properties of the 
quantum regime can be described.

\subsection{Experimental possibilities}
The measurement statistics described here should be obtainable by carefully
evaluating the data obtained in any quantum nondemolition measurement
followed by a measurement of field coherence, e.g. by homodyne detection.
It is important, however,
to keep track of the correlation between the measurement result $n_m$
and the corresponding average results of the field measurements 
$\langle \hat{a} \rangle_f(n_m)$. This requires some amount of time resolution,
for example in the form of light field pulses or perhaps of solitons in 
fibers \cite{Fri92}.
Unfortunately, it is extremely difficult to realize quantum nondemolition
measurements of high resolution in the optical regime. The experimental
results cited here \cite{Lev86,Fri92} are still well in the classical 
regime of $\delta\! n > 1$. Possibly, a realization based on the interaction
of single atoms with a microwave mode \cite{Bru90,Hol91} might be more
promising. In particular, the use of a variable number of single probe atom 
passed through the cavity should allow a particulatily reliable variation of 
the photon number resolution parameter $\delta\! n$.

The challenge presented by the aspects of quantum theory discussed above
is to obtain sufficient control of quantum coherence to explore the
properties at the very limit of quantum mechanical uncertainty. The effects
observed in this regime should then help to illustrate 
the quantum mechanical properties utilized for quantum computation,
quantum communication, and other aspects of quantum information \cite{Hof99}. 
The continuous transition from the classical aspects of optical coherence
to the quantum properties of the light field can also serve as a tool 
to pinpoint the technological requirements for more complex 
implementations of quantum optical devices.

\section*{Acknowledgements}
The Author would like to acknowledge support from the Japanese Society for
the Promotion of Science, JSPS.



\end{document}